\documentclass[%
 reprint,
 superscriptaddress,
 aps,
 pra,
]{revtex4-2}
\usepackage{amsmath}
\usepackage{amssymb}
\usepackage{amsthm}
\usepackage{graphicx}
\usepackage{dcolumn}
\usepackage{bm}

\usepackage[english]{babel}
\makeatletter
\let\ORIbbl@fixname\bbl@fixname
\def\bbl@fixname#1{%
  \@ifundefined{languagealias@\expandafter\string#1}
    {\ORIbbl@fixname#1}
    {\edef\languagename{\@nameuse{languagealias@#1}}}%
}
\newcommand{\definelanguagealias}[2]{%
  \@namedef{languagealias@#1}{#2}%
}
\makeatother
\definelanguagealias{en}{english}
\definelanguagealias{eng}{english}

\usepackage{enumitem} 
\usepackage{braket} 
\usepackage{mathtools} 
\usepackage[dvipsnames]{xcolor}
\usepackage{comment} 
\usepackage{physics} 

\usepackage[font=small, justification=Justified]{caption}
\usepackage{subcaption}

\renewcommand{\equiv}{\coloneqq}

\newcommand{\intP}{\int_{\mathcal{P}(\mathcal{H})} \!\!\!\!\!\!\!\!\!}

\newcommand{\PH}{\mathcal{P}(\mathcal{H})}
\newcommand{\Poiss}[2]{\left\{ #1, #2\right\}_{\mathrm{PB}}}
\usepackage{amsthm}
\theoremstyle{definition}

\definecolor{S_Blue}{RGB}{0,135,252}
\definecolor{S_Red}{RGB}{214,13,63}
\definecolor{Blue}{RGB}{47,89,151}
\definecolor{S_Grey}{RGB}{150,150,158}
\definecolor{S_Yel}{RGB}{255,204,0}
\definecolor{S_Green}{RGB}{102,204,0}
\definecolor{S_Brown}{RGB}{154,41,41}

\usepackage{mathrsfs} 

\usepackage{stmaryrd} 

\usepackage{pgffor}
\usepackage{ifthen}
\usepackage{placeins}
\usepackage[unicode=true]{hyperref} 

\begin{document}


\title{Non-equilibrium theory of projected ensembles}

 \author{Fabio Anza}
 \email{fanza@umbc.edu}
 \author{Cameron Hahn}
 \email{chahn3@umbc.edu}
  \affiliation{
 Department of Physics, Cybersecurity Institute, Quantum Science Institute, University of Maryland, Baltimore County, Baltimore, Maryland, US
 }

\date{\today}

\begin{abstract}
Projected ensembles---the collections of conditional pure states induced on a system by measuring an entangled environment---have become central objects in quantum science, underlying deep thermalization, quantum state designs, the emergence of classicality, and exhibiting interesting phase transitions. Here we develop a general and exact theory of their dynamics, yielding a systematic approach to their dynamics and equilibrium and non-equilibrium stationary states. We derive an exact continuity equation on quantum state space for the projected ensemble, together with microscopic kinetic equations and analytical expressions for the probability flux and source terms generated by the system–environment interaction. The resulting dynamics admits a classical representation: isolated systems obey Hamiltonian transport and Liouville’s theorem, while open systems are described by a kinetic theory of probability transport. The stationary continuity equation provides a general characterization of equilibrium and non-equilibrium stationary projected ensembles, which can be analyzed through the method of characteristics. The theory provides an analytical framework for studying the emergence, structure, and timescales of equilibrium and non-equilibrium stationary projected ensembles, with applications ranging from deep thermalization and the emergence of classicality to random quantum-state generation and quantum device benchmarking.
\end{abstract}

\maketitle

\section*{Introduction.} 

Understanding the interplay between randomness and determinism in unitary evolution in many-body quantum systems underlies many foundational and applied challenges in modern quantum science: from the measurement problem and the emergence of classicality, to black-hole physics and quantum device benchmarking. The study of random quantum states has recently emerged as a frontier of interest, aimed at understanding and controlling ensembles of random pure states. The emergence of ensembles clustered around pointer states, predicted to underlie the emergence of classicality \cite{touil_branching_2024}, was experimentally verified in superconducting quantum circuits\cite{zhu_observation_2025}. Recent studies of Hilbert space ergodicity~\cite{markMaximumEntropyPrinciple2024,pilatowsky-cameo_hilbert-space_2024} and of the emergence of Gibbs ensembles of pure states~\cite{Anza2022GeometricQuantumThermodynamics, hoExactEmergentQuantum2022, ippolitiSolvableModelDeep2022,goldsteinUniversalProbabilityDistribution2016,goldsteinDistributionWaveFunction2006}, dubbed \emph{deep thermalization}, have expanded our understanding of how statistical mechanics emerges from within a unitary dynamics. On the applied side, random quantum states are increasingly viewed as a resource~\cite{zhangHolographicDeepThermalization2025, chakrabortyFastComputationalDeep2025,bejanMatchgateCircuitsDeeply2025}: state preparation~\cite{magann_randomized_2023,choiPreparingRandomStates2023,cotlerEmergentQuantumState2023}, benchmarking of quantum devices~\cite{cross_validating_2019}, certification of quantum advantage~\cite{arute_quantum_2019}, cryptography~\cite{pirandola_advances_2020}, and quantum learning~\cite{elben_randomized_2022} all make essential use of random states.

While ensembles of pure states can arise in several physically interesting ways~\cite{santiniSemiclassicalQuantumTrajectories2025,boormanDiagnosticsEntanglementDynamics2022,michaelbuchholdEffectiveTheoryMeasurementInduced2021,gullansDynamicalPurificationPhase2020}, \emph{projected ensembles} have recently attracted particular attention. Measuring an environment entangled with a system of interest leaves the system in a pure state conditioned on the measurement outcome; the collection of these conditional states, weighted by the corresponding outcome probabilities, forms the projected ensemble. In chaotic many-body systems, projected ensembles have been shown to approach universal forms associated with deep thermalization, including the Haar measure and quantum $t$-designs~\cite{ambainis_quantum_2007,hoExactEmergentQuantum2022,ippolitiSolvableModelDeep2022,markMaximumEntropyPrinciple2024}, as well as finite-temperature generalizations~\cite{goldsteinUniversalProbabilityDistribution2016,Anza2022GeometricQuantumThermodynamics}. More broadly, their statistics have also emerged as probes of qualitatively distinct dynamical regimes and of phase transitions in monitored and many-body quantum systems~\cite{boormanDiagnosticsEntanglementDynamics2022,michaelbuchholdEffectiveTheoryMeasurementInduced2021,gullansDynamicalPurificationPhase2020,minatoFateMeasurementInducedPhase2022}. 


In this work we tackle the non-equilibrium phenomenology of projected ensembles. We develop a
general and exact theory of the dynamics of projected ensembles by exploiting the symplectic
structure of quantum state space, in synergy with the geometric formulation of quantum mechanics~\cite{STROCCHI1966ComplexCoordinatesQuantum,Kibble1979GeometrizationQuantumMechanics,Heslot1985QuantumMechanicsClassical,Ashtekar1999GeometricalFormulationQuantum,Brody2001GeometricQuantumMechanics,Bengtsson2017GeometryQuantumStates,Anza2021DensityMatricesGeometric,Anza2022GeometricQuantumThermodynamics}. This allows us to formulate projected-ensemble dynamics as a problem of probability transport on the manifold of pure states and tackle several crucial questions: how probability moves across
quantum state space, how the interaction with the environment redistributes it, and what equilibrium and non-equilibrium stationary structures can emerge from this dynamics. 

We derive an exact continuity equation for the ensemble, together with microscopic expressions for its probability flux and sink/source terms, and we show how the resulting transport theory characterizes stationary ensembles analytically through the method of characteristics. The resulting theory connects microscopic unitary dynamics directly to ensemble-level transport, and provides a concrete analytical toolbox for the study of non-equilibrium properties of projected ensembles.

The paper is organized as follows. In Section \ref{sec:GQM} we give a brief summary of
GQM, and of the tools introduced in \cite{Anza2021DensityMatricesGeometric,Anza2022GeometricQuantumThermodynamics} to describe ensembles.
In Section \ref{sec:IT} we provide our first result: a microscopic derivation of the continuity equation
for the transport of probability across state space in quantum systems, together with the appropriate microscopic definitions
for the relevant phenomenological quantities: \emph{the probability flux and the probability sink/sources}, together
with their kinetic interpretation. In Section \ref{sec:MOC} we show how the method of
characteristics turns the stationary continuity equation into an ordinary differential equation along the flow,
giving a general route to stationary geometric quantum states. Sections \ref{sec:IQS} and \ref{sec:DYN} deal with the kinetic aspects underlying the
continuity equation and give the second result: the set of microscopic equations regulating the transport of
probability, first for isolated systems, where the flux reduces to a Hamiltonian flow and a Liouville theorem holds,
and then for a system coupled to a finite environment, where explicit expressions for $J_t$ and $\sigma_t$ follow from
a set of coupled equations for interacting probability carriers.  Sections \ref{sec:EXAMPLES2} and \ref{sec:SPINSTAR}
put the framework to work on two concrete models: a qubit in a Caldeira-Leggett model treated via Langevin dynamics, where a Fokker-Planck equation for the ensemble is derived and shown to be equivalent to the continuity equation; and a qubit in a spin-star environment, where the stationary projected ensemble is obtained analytically and checked against exact diagonalization. In Section \ref{sec:FINAL} we draw some conclusions.



\section{Geometric Quantum Mechanics}
\label{sec:GQM}

References
\cite{STROCCHI1966ComplexCoordinatesQuantum,Mielnik1968GeometryQuantumStates,Kibble1979GeometrizationQuantumMechanics,Heslot1985QuantumMechanicsClassical,Page1987GeometricalDescriptionBerrys,Anandan1990GeometryQuantumEvolution,Gibbons1992TypicalStatesDensity,Ashtekar1999GeometricalFormulationQuantum,Brody2001GeometricQuantumMechanics,Bengtsson2017GeometryQuantumStates,Carinena2007GeometrizationQuantumMechanics,Chruscinski2006GeometricAspectsQuantum,Marmo2010GeometricalDescriptionQuantum,Avron2020ElementaryIntroductionGeometry,Pastorello2015GeometricHamiltonianFormulation,Pastorello2015GeometricHamiltonianDescription,Pastorello2016GeometricHamiltonianQuantum} give a comprehensive introduction to GQM. Here, we briefly summarize only the elements we need, working with Hilbert spaces $\mathcal{H}$ of finite dimension $D$. For the details of the derivations, we send the reader to the literature cited above. Given an arbitrary basis $\{\ket{c_{n}}\}_{n=0}^{D-1}$ a pure state is parameterized by $D$ complex homogeneous coordinates $Z=\{Z^{n}\}_{n=0}^{D-1}$, up to normalization and an overall phase:
\begin{equation}
\ket{\psi} = \sum_{n=0}^{D-1}Z^{n}\ket{c_{n}} \label{eq:homogeneous-coords}
\end{equation}
Here, we will always use upper indices to identify different coordinates of the same point and lower indices to identify different points. The Hilbert space formulation is redundant since the multiplication of $Z$ by a complex number $\lambda \in \mathbb{C}$ changes the vector representation but not the physical state. Therefore, $Z\in \mathbb{C}^{D}$, $Z\sim\lambda Z$, with $\lambda\in \mathbb{C}\setminus\{ 0 \}$. This equivalence relation means pure states of a quantum system are points in the complex projective space $\mathcal{P}(\mathcal{H})\sim \mathbb{C}P^{D-1}$. We refer to $\mathcal{P}(\mathcal{H})$ as \textit{quantum state space} or as the \textit{manifold of quantum states}. On $\mathcal{P}(\mathcal{H})$, one can always use probability-phase coordinates $Z^{n}=\sqrt{ p^{n} }e^{i\phi^{n}}$ which as we see now, play a particularly important role.

It is well known that $\mathcal{P}(\mathcal{H})$ is a K\"ahler manifold \cite{Bengtsson2017GeometryQuantumStates}, with two important geometric structures intertwined with each other. First, it has a preferred metric $g_{FS}$ - the \textit{Fubini-Study metric} \cite{Bengtsson2017GeometryQuantumStates} - and an associated volume form $dV_{FS}$ that is coordinate-independent and invariant under unitary transformations. The general expression of $dV_{FS}$ is known \cite{Bengtsson2017GeometryQuantumStates} but, for our purposes, it is sufficient to give its explicit form in probability-phase coordinate
\begin{equation}
dV_{FS} = \prod_{n=1}^{D-1} \frac{dp_{n}d\phi_{n}}{2}~.\nonumber
\end{equation}
Second, it has a symplectic structure, seen from the fact that probability-phase coordinates are canonically conjugated. This enables us to define ``Poisson Brackets" and accurately refer to $\mathcal{P}(\mathcal{H})$ as the \textit{quantum state space}. The symplectic structure enables us to describe the evolution of the probability-phase coordinates of a quantum state using Hamilton's equations of motion

\begin{align}\label{eq:HAM_EOM}
    &\frac{dp^n}{dt} = \frac{1}{\hbar}\frac{\partial E}{\partial \phi^n}~, &&
    \frac{d\phi^n}{dt} = -\frac{1}{\hbar}\frac{\partial E}{\partial p^n}
\end{align}

The geometric framework makes it very natural to view a quantum state as a functional encoding that associates expectation values to observables in a linear way, paralleling the $C^{*}$-algebra formulation of quantum mechanics \cite{Strocchi2008IntroductionMathematicalStructure}. This linearity suggests the use of measure-theoretic probability, paralleling the use of ensembles in classical statistical mechanics.
\begin{figure*}
    \centering
    \includegraphics[width=0.95\textwidth]{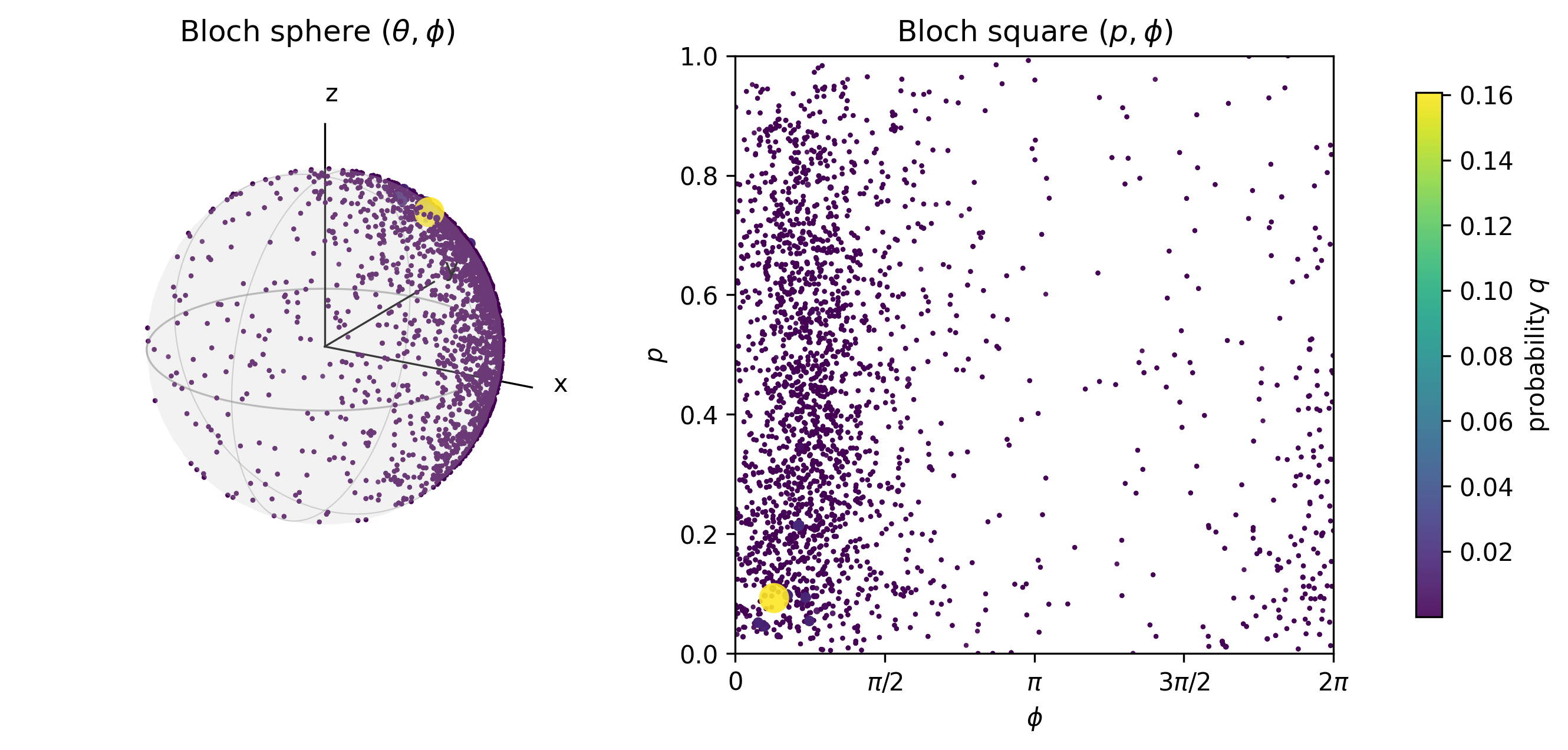}
    \caption{An example of a geometric quantum state for a qubit, $D=2$. Each point is one element $\Gamma_\alpha$ of the ensemble, carrying a probability mass $x_\alpha$ (color scale). Left: A representation of the ensemble on the Bloch sphere. Right: the same ensemble in canonically conjugated probability-phase coordinates, in which the state of the qubit is parameterized as $\ket{\psi} = \sqrt{1-p}\,\ket{0} + \sqrt{p}\,e^{i\phi}\ket{1}$ and the state space becomes the ``Bloch square'' $[0,1]\times[0,2\pi)$, with the poles $p=0,1$ mapping to its lower and upper edges.}
    \label{fig:gqs_dynamics}
\end{figure*}
To consider ensembles in the manifold of quantum states, we turn $\mathcal{P}(\mathcal{H})$ into a probability space $\mathbb{P}=(\mathcal{P}(\mathcal{H}), \mathcal{B}(\mathcal{H}),\mu)$, where $\mathcal{B}(\mathcal{H})$ is the Borel $\sigma$-algebra of open sets on $\mathcal{P}(\mathcal{H})$, and $\mu$ is a normalized measure $\mu(\mathcal{P}(\mathcal{H}))=1$ on $\mathcal{B}(\mathcal{H})$. The simplest probability measure is the normalized version of the Fubini-Study volume measure $d\nu_{FS}=\frac{dV_{FS}}{V_{D-1}}$, where $V_{D-1}=\frac{\pi^{D-1}}{(D-1)!}$ is the Fubini-Study volume of $\mathcal{P}(\mathcal{H})$. By its invariance under unitary transformations, this is the basic definition of the uniform distribution in $\mathcal{P}(\mathcal{H})$. In general, the measure $\mu$ is not uniform. When $\mu$ is absolutely continuous with respect to $V_{FS}$, Radon-Nikodym's theorem guarantees the existence of a non-negative probability density function $q(Z=z)$ such that $\mu(d\nu_{FS}^{z}) = q(z)dV_{FS}^{z}$. Then $\mu(d\nu^{z})$ is the infinitesimal probability that the system's state $Z$ will belong to an infinitesimal volume $dV_{FS}^{z}$ centered on $z\in \mathcal{P}(\mathcal{H})$. In cases in which $\mu_t$ is not absolutely continuous with respect to $dV_{FS}$, we can still use the density $q(z)$ but we have to remember that it will be a distribution (generalized function). This allows us to retain the standard calculus-based notation while dealing with singular measures like Dirac's measure, for which the density is the well-known Dirac's delta distribution.

We call both $\mu$ and $q$ the system's Geometric Quantum State (GQS) or ensemble, as in GQM, these objects are identical. In this geometric description, the quantumness of the system is encoded in the non-trivial geometry of the sampling space: the state space of a quantum system. 

For much of this work, we concern ourselves with discrete ensembles of pure states. The GQS of such an ensemble can be expressed as
\begin{equation}
\mu = \sum_{\alpha=0}^{N} x_\alpha \delta_{Z(\chi_\alpha)}~,\label{eq:gqs_opn}
\end{equation}
where $N$ is the number of pure states contributing to the GQS, $x_\alpha$ is the probability associated with the pure state $\ket{\chi_\alpha}$ and $\delta_{Z(\chi_\alpha)}$ picks out the point in $\mathbb{C}P^{D-1}$ corresponding to the pure state $\ket{\chi_\alpha}$ through the homogeneous coordinates $Z(\chi_\alpha)$. A visual example of a geometric quantum state is the finite ensemble represented in Figure \ref{fig:gqs_dynamics}.

\section{Probability Transport: Continuity equation}
\label{sec:IT}

So far, we have summarized previous results about GQM. We now build on them 
and, by bringing in the dynamics of the system, we derive 
a continuity equation which dictates how the geometric quantum state of an open
quantum system out-of-equilibrium evolves, under very general assumptions. This is 
the fundamental kinetic equation governing how the probability of finding the state 
of the system in a region of $\PH$ changes as a result of its interaction with an environment. 
Throughout this section we will try to maintain a fairly general language, to emphasize
how the treatment applies to quantum systems under very general assumptions. However, 
for concrete examples, both numerical and analytical, we will always refer to the case of a qubit.

\subsection*{General treatment}
The following treatment, and its results, pertains quantum systems which are finite-dimensional, 
and interact with finite-dimensional environments but are, otherwise, arbitrary. 

\begin{figure*}
    \centering
    \includegraphics[width=0.95\textwidth]{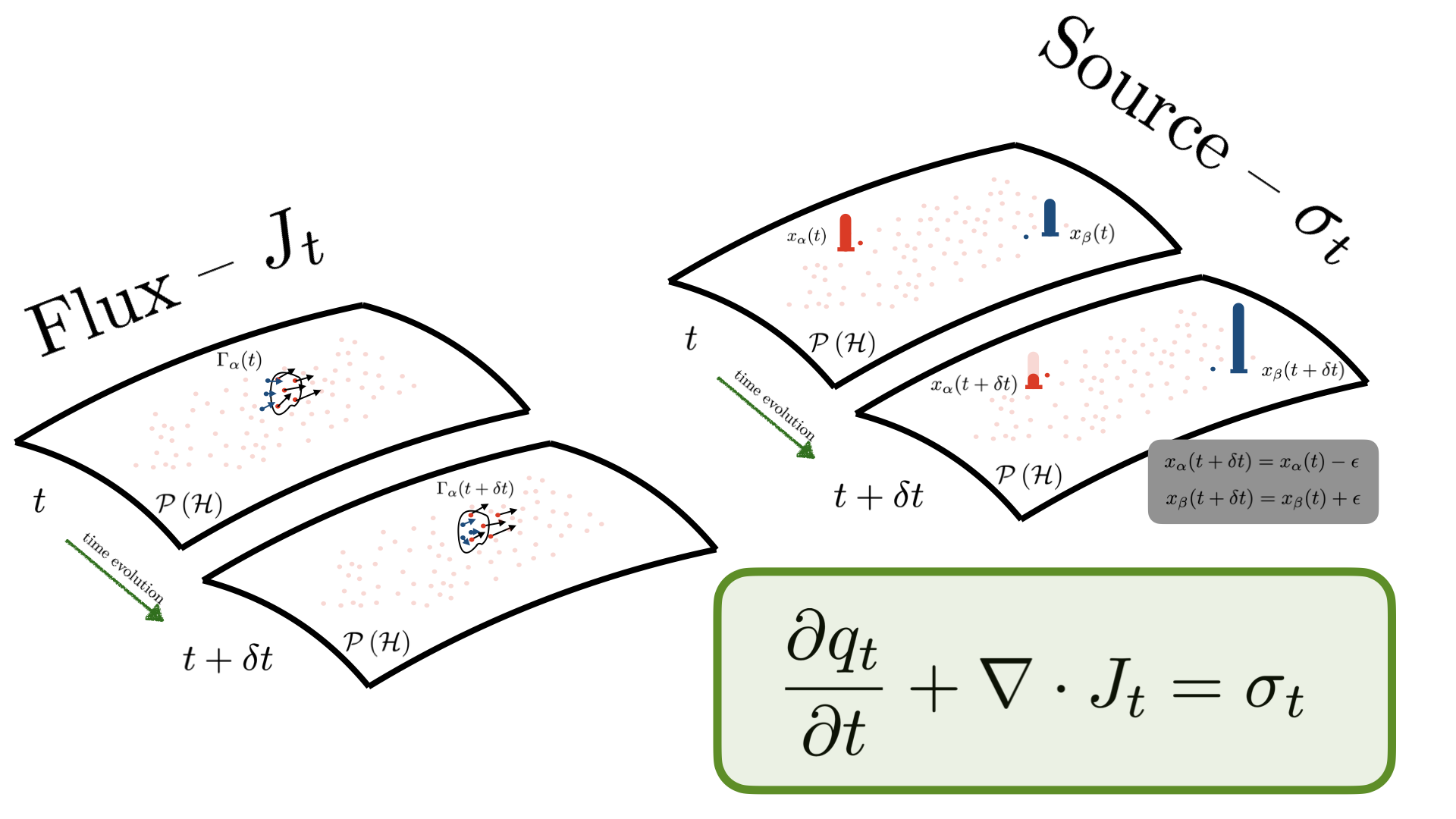}
    \caption{A diagrammatic view of the continuity equtaion. Left: Kinetic interpretation of the probability flux $J_t$. If we look at a small region of the quantum state space,  due to the underlying dynamics we see that probability is locally conserved 
	in the sense that there is a certain number of points which enters and leaves this region. As a 
	result of this local process, probability is moved around and can concentrate in a certain region 
	or get scrambled across the quantum state space. Right: Kinetic interpretation of the sink/source term $\sigma_t$. Even if the points do not move around
	the quantum state space $\dot{\Gamma}_\alpha=0$, the term $\sigma_t$ allows the exchange of probability 
	between different regions. Note that this can be a non-local effect in state space, as depicted above. In the 
	example above	we have two states $\Gamma_\alpha$ and $\Gamma_\beta$ fixed in time, but exchanging 
	a certain amount $\epsilon$ of probability, thus moving probability from one region of the state space to 
	another one: $x_\beta(t+\delta t) = x_\beta(t)+\epsilon$, $x_\alpha(t+\delta t) 	= x_\alpha(t) - \epsilon$.}
    \label{fig:continuitydiagram}
\end{figure*}

A geometric quantum state is specified by two sets of quantities: $\left\{x_\alpha\right\}_{\alpha=0}^{N}$
and $\left\{ \Gamma_\alpha\right\}_{\alpha=0}^{N}$ which is a short-hand notation for 
$\Gamma_\alpha = Z(\ket{\chi_\alpha})$ with $Z$ being the homogeneous coordinates defined by~\eqref{eq:homogeneous-coords}. The first one is a classical probability distribution across the pure states $\ket{\chi_\alpha}$ which are members of the ensemble. In this way, $x_\alpha$ determines the weight of each pure state in the overall statistical mixture. The pure states themselves are described in the ensemble via their homogeneous coordinates $\Gamma_\alpha$ which live in the complex projective space: $\Gamma_\alpha \in \mathcal{P}(\mathcal{H}_S) = \mathbb{C}P^{D-1}$ with $D$ the dimension of the system Hilbert space,
which corresponds to the ket $\ket{\chi_\alpha} \in \mathcal{H}_S$. Since these $\Gamma_\alpha(t)$ are points moving 
in a state space, we can think of them as particles on a classical phase space, 
which interact in a non-trivial way. These are the \emph{``carriers of probability''}, in the sense that each of 
these particles carries a probability mass $x_\alpha(t)$ that the system will be found in $\Gamma_\alpha(t)$.
An example of the geometric quantum state of a qubit is given in Figure \ref{fig:gqs_dynamics}. Since the total amount 
of probability has to be preserved, $\mu_t(\PH) = \sum_\alpha x_\alpha(t) =1$,
the geometric quantum state $\mu_t$ must satisfy a continuity equation. Conceptually, this is 
the starting point of virtually all transport theories, which deal with quantities that are globally conserved, but that are moved around as a result of the underlying microscopic dynamics.

Thus, in the dynamics of a geometric quantum state we identify two different terms, arising from the 
time-evolution of these two sets of quantities. There are two different ways in which a geometric quantum state can change in time. First, the amount of probability mass carried by points can change: $\dot{x}_\alpha \neq 0$, see Figure \ref{fig:continuitydiagram}. Second, there is
the movement of each point: $\dot{\Gamma}_\alpha \neq 0$, see Figure \ref{fig:continuitydiagram}. By summing each term over all points, and then summing the two terms, we get the following evolution equation for the geometric quantum state:
\begin{equation}
\dot{\mu}_t = \sum_\alpha \dot{x}_\alpha \delta_{\Gamma_\alpha} - x_\alpha \dot{\Gamma}_{\alpha} \delta^{'}_{\Gamma_\alpha} ~.\label{eq:cont}
\end{equation}
Let's start by analyzing the second term. $\delta^{'}$ is the distributional derivative 
of the Dirac measure $\delta_{\Gamma_\alpha}$ and $\dot{\Gamma}_\alpha(t)$ is 
the velocity vector, in state space, of each point $\Gamma_\alpha(t)$. Thus, this term 
is the covariant divergence of a velocity field or, in other words, a flux term in a continuity equation. 
Indeed, the whole term can be explicitly written as the divergence,
\begin{equation}
\sum_\alpha x_\alpha \delta^{'}_{\Gamma_\alpha} \dot{\Gamma}_{\alpha} = \nabla \cdot J_t~, \nonumber
\end{equation}
where we have identified the \emph{Probability Flux} $J_t$ as the ``single-state'' flux $J_\alpha(t)$, averaged
with the probability mass it is carrying $x_\alpha(t)$:
\begin{subequations}\label{eq:flux}
\begin{align}
&J_t = \sum_\alpha x_\alpha(t)J_\alpha(t)~,\\
&J_\alpha(t) =  \delta_{\Gamma_\alpha(t)} \dot{\Gamma}_{\alpha}(t)~.
\end{align}
\end{subequations}
We now look at the first term in the right-hand side of Eq.~\eqref{eq:cont}: $\sum_\alpha \dot{x}_\alpha(t) \delta_{\Gamma_\alpha(t)}$.
We recognize a source/sink term, which is independent on the underlying movement of
the points in the quantum state space. This identifies the other functional term in our continuity 
equation: \emph{sinks and sources of probability} $\sigma_t$:
\begin{equation}
\sigma_t = \sum_{\alpha=0}^{d_E-1} \dot{x}_\alpha(t) \delta_{\Gamma_\alpha(t)}~.\label{eq:sigma}\nonumber
\end{equation}
Eventually, we obtain the following continuity equation for the geometric quantum state of a finite-dimensional
quantum system interacting with a finite-dimensional environment:
\begin{equation}
\dot{\mu}_t + \nabla \cdot J_t = \sigma_t\label{eq:continuity}
\end{equation}
This is our first and most important result. It is the central tool which describes, in a phenomenologically
meaningful and analytically tractable way, how probability dynamically spreads around the state space.

Before we move on and use the equation to study the flow of probability in some concrete
systems, a few comments about general validity of the equation, and the physical interpretation of each of
its terms are in order.

\subsection*{Discussion and interpretation}

First, the derivation presented here pertains ensembles of finite-dimensional quantum systems made up by a finite number of points $N_P$. However, generalization to countably-infinite or uncountably-infinite dimensional
environment is straightforward. In the first case, a simple limit $N_P \to \infty$ suffices. In most cases the limit goes through the derivation, leaving it unaltered. However, singular limits are possible and one must always be mindful
of the physical meaning of the limit that is being enforced. In the second case, which pertains the case in which we have continuous ensembles, only minor adjustments are needed. The core aspects of the derivation hold, leaving the final form of the continuity equation unaltered. In summary, the only difference between the three cases of finite, countably-infinite or uncountably-infinite
ensembles will be that the support of the geometric quantum state will go from being a set of dimension zero (finite number
of points), to a more generic set which, in some cases, can even have fractal dimension. For a thorough analysis of these situations we send the interested reader to Ref.\cite{Anza2022QuantumInformationDimension}.

Second, an aspect that we deem quite interesting here is that the proposed theoretical framework allows us to talk 
about probability as a localized quantity, which is carried around in quantum state space. This is in analogy 
with all classical theories of transport of properties, such as mass or charge, which are carried around by 
particles, represented as points wandering in a classical phase-space.

Third, the derivation was performed using the generic measure $\mu_t$. Thus, it holds also for a smooth probability density $q_t(Z)$, which exists when the Radon-Nikodym derivative $\frac{d\mu_t}{d\nu_{FS}}$ of $\mu_t$ with respect to the uniform Fubini-Study measure exists.

Fourth, to strengthen the physical intuition about the theoretical derivation, we now argue that the physical interpretation 
of the flux $J_t$ and source/sink $\sigma_t$ terms are essentially the same as in other transport theories. Indeed, the 
property that is carried around is probability mass. We now describe them explicitly, using 
their kinetic interpretation, drawn in Figure \ref{fig:continuitydiagram}. 

\paragraph*{$J_t$ is a probability flux,} associated to an overall conserved quantity, probability mass, localized in its carriers $\Gamma_\alpha$, which 
is moved around the quantum state space. To emphasize this point, we look at the situation in which $\dot{x}_\alpha = 0$. 
In this case, $\sigma_t=0$ and the probability carried around by the points in state space is a constant of motion 
($x_\alpha(t)= x_\alpha(t_0)$). Thus, the analogy with a typical model of point particles carrying around a physical property, 
like charge or mass, becomes exact. As the carriers move around, the probability is dispersed across the quantum 
state space in a way that guarantees its local preservation. This means the spread of probability can be described 
with a continuity equation, which equates the local time derivative of the probability distribution with the divergence of the 
local flux of probability.

\paragraph*{$\sigma_t$ is a source term,} with the usual interpretation. Indeed, first, we note that it
can not be written as a divergence term on the quantum state space. Second, we look at the simpler dynamics in 
which the position of the points in $\PH$ does not change: $\dot{\Gamma}_\alpha = 0$. In this case the support 
of the distribution is fixed and the particles don't move around $\Gamma_\alpha(t)=\Gamma_\alpha(t_0)$. However, 
there can still be a non-trivial dynamics, due to $\dot{x}_\alpha(t)\neq 0$. Depending on the actual support of 
the distribution, as depicted in Figure \ref{fig:continuitydiagram}, the time-evolution generated by this term is generally non-local in the 
quantum state space. Thus, $\sigma_t$ possesses all the hallmarks of the standard sinks and sources terms in 
transport theories: it is not associated with particles moving around the state space and it can not be rewritten 
as a divergence term, while still affecting the time-derivative of the quantity that is being transported---probability mass.

\section{Solving the Continuity Equation: Method of Characteristics}
\label{sec:MOC}

Sections \ref{sec:IQS} and \ref{sec:DYN} derived microscopic expressions for the flux and source terms in closed and open configurations. The question still remains: how can we use this framework to obtain practical information about the behavior and dynamics for systems of interest? One of the natural uses is finding stationary ensembles under particular model dynamics. To facilitate such a solution, we utilize the method of characteristics developed for solving partial differential equations.

In the case of a stationary solution, $\dot{\mu}_t = 0$ reducing the continuity equation to the form
\begin{equation}
    \nabla\cdot J = \sigma \nonumber
\end{equation}
We consider the flux to have the form $J = qv$ analogous to \eqref{eq:Ohmic-flux} with the geometric state $q$ and the flow $v$ being parameterized by the homogeneous coordinates $Z$. In general the flow $v$ does not need to be Hamiltonian meaning that $\nabla\cdot v \ne 0$ and the continuity equation remains as
\begin{equation}
    v(Z)\cdot\nabla q(Z) + q(Z)\nabla\cdot v(Z) = \sigma(Z) \nonumber
\end{equation}
To solve this PDE, we can employ the method of characteristics by introducing the characteristic parameter $s$ and writing the total derivative of the geometric quantum state with respect to the characteristic parameter as
\begin{equation}
    \frac{dq}{ds} = \sum_n\frac{\partial q}{\partial Z^n}\frac{dZ^n}{ds} \nonumber
\end{equation}
The derivative of the $n$th coordinate with respect to the characteristic parameter can be identified as the $n$th component of the flow $\frac{dZ^n}{ds} = v^n(Z(s))$, allowing the total derivative of the state to be represented by
\begin{equation}
    \frac{dq}{ds} = v\cdot\nabla q \nonumber
\end{equation}
which reduces the continuity equation to the following ODE:
\begin{equation}
    \frac{dq}{ds} + (\nabla\cdot v)(Z(s))q = \sigma(Z(s)) \nonumber
\end{equation}
Providing that $\sigma(Z(s))$ and $\nabla\cdot v$ are independent of the state $q$, this is a first order ODE which can be solved via an integrating factor $e^{\Lambda(s)}$ with $\Lambda(s) = \int_0^S(\nabla\cdot v)(Z(s'))\, ds'$ such that the general stationary solution is expressed as
\begin{equation}
    q(Z(s)) = e^{-\Lambda(s)}\left[ q(Z(0)) + \int_0^se^{\Lambda(s')}\sigma(Z(s'))\, ds' \right] \nonumber
\end{equation}
Importantly, $q(Z(0))$ serves as an initial condition with respect to the characteristic curve, not with respect to the particular evolution being studied. We can eliminate $s$ from this expression for the stationary state by using the definition we imposed for the flow
\begin{equation}
    \frac{dZ^n}{ds} = v^n(Z(s)) \nonumber
\end{equation}
by solving and inverting the $n$ resulting ODEs.

It is worth recording the simplest special case, which we will use later. Suppose the dynamics is Hamiltonian,
$\nabla \cdot v = 0$, and the sink/source term vanishes, $\sigma = 0$. Then $\Lambda(s) = 0$ and the solution
above collapses to
\begin{equation}
    q(Z(s)) = q(Z(0))~,\label{eq:stat-level-set}\nonumber
\end{equation}
that is, the geometric quantum state is constant along each characteristic curve. 


For a Hamiltonian flow, the characteristics are the integral curves of $v_H$, and conservation of energy implies that each characteristic is contained in a level set of $E(Z)$. Equation~\eqref{eq:stat-level-set} therefore states, in full generality, that a stationary geometric quantum state is constant along Hamiltonian trajectories,
\begin{equation}
\Poiss{q}{E}=0~.\nonumber
\end{equation}
If additional conserved quantities are present, the stationary state may depend on them as well. In the single-qubit case relevant to Section~\ref{sec:SPINSTAR}, regular connected energy level sets are one-dimensional and coincide with the Hamiltonian trajectories. In that case the stationary solution reduces to
\begin{equation}
q(Z)=f(E(Z))~,\label{eq:stat-energy}
\end{equation}
with $f$ fixed by the initial condition. This is the geometric counterpart of the familiar statement that stationary classical distributions are functions of the conserved quantities, and it is the form we will encounter in Section \ref{sec:SPINSTAR}.

\section{Isolated-system dynamics and Liouville's theorem}\label{sec:IQS}

As a first use-case, here we describe the flow of probability when our system of interest is
isolated. In this case both $J_t$ and $\sigma_t$ can be characterized analytically in full generality. To ease the exposition of the results, in this subsection we will use the density $q_t$ such that $\mu_t(dV_{FS})=q_t(Z)dV_{FS}$, 
where $q_t$ is a distribution.

A clarification on terminology is in order. By ``isolated'' we mean here that every element of the ensemble is carried by a single Hamiltonian flow on $\mathbb{C}P^{D-1}$, generated through the symplectic form~\eqref{eq:HAM_EOM} by the energy function $E(Z) = \bra{\psi(Z)}H\ket{\psi(Z)}$ of a fixed Hamiltonian $H$, and that the statistical weight attached to each element is a constant of motion.

The elementary case is unitary evolution of a single pure state, for which the ensemble is a Dirac measure $\delta_{\chi(t)}$, where  $\ket{\chi(t)} = U(t)\ket{\chi(0)}$, so the ensemble remains a Dirac measure of unit weight and the entire dynamics reduces to the motion of the support. A finite ensemble $\left\{ x_\alpha, \ket{\chi_\alpha}\right\}_\alpha$ is a classical mixture of such cases, and since the same $U(t)$ acts on every element in an isolated system, the result carries over additively:
\begin{equation}
    q_t(Z) = \sum_\alpha x_\alpha\,\delta_{\Gamma_\alpha(t)} \nonumber
\end{equation}
where $\Gamma_\alpha$ are the state space points on which the ensemble has non-vanishing support, corresponding to $\ket{\chi_\alpha(t)}$. Here we write the support as discrete for readability, though everything we say here holds for arbitrary measures.

Two features are relevant here. First, $x_\alpha(t)=x_\alpha(0)$, leading to $\dot{x}_\alpha=0$. In turn, this leads to $\sigma_t=0$:
This confirms that in order to have $\sigma_t \neq 0$ we need a dissipative dynamics, resulting from the interaction of our system with an environment. Second, $\ket{\chi_\alpha(t)}=U(t)\ket{\chi_\alpha(0)}$. This
means all elements $\Gamma_\alpha(t)$ of the support of $q_t$ evolve rigidly, according to the same equation. Indeed, starting from the 
definition of the flux in Eq.~\eqref{eq:flux}, we can explicitly write the form of the flux $J_t$, 
using Hamilton's equations of motion. Writing $\Gamma_\alpha$ in canonical coordinates and with the following vector notation $\Gamma_\alpha = (\vec{p}(\Gamma_\alpha),\vec{\phi}(\Gamma_\alpha)) = (\vec{p}_\alpha,\vec{\phi}_\alpha)$, 
we have 
\begin{align*}
\dot{\Gamma}_\alpha & = \left( \frac{d\vec{p}_\alpha}{dt}, \frac{d\vec{\phi}_\alpha}{dt}\right) \nonumber \\
& = \left.\left( \frac{1}{\hbar}\frac{\partial E}{\partial \vec{\phi}}\right\vert_{\Gamma_\alpha}, \left.-\frac{1}{\hbar}\frac{\partial E}{\partial \vec{p}}\right\vert_{\Gamma_\alpha}  \right) = v_H(\vec{p}_\alpha,\vec{\phi}_\alpha)
\end{align*}
Note how the flow, $v_H$, does not depend on the index $\alpha$: all 
$\Gamma_\alpha$ evolve according to the same (Hamiltonian) flow. Inserting this into the definition of $J_t$ 
we obtain
\begin{equation}
J_t = q_t v_H~, \label{eq:Ohmic-flux}
\end{equation}
leading to the following continuity equation:
\begin{equation}
\frac{\partial q_t}{\partial t} = - v_H \cdot \nabla q_t - q_t \, \nabla \cdot v_H \nonumber
\end{equation}
This can be further simplified by remembering that the Hamiltonian vector field is divergence-free, $\nabla \cdot v_H = 0$.
For a discussion on this (and more) see Ref.\cite{Marsden1999IntroductionMechanicsSymmetry}. This leads us to the final form of the continuity equation of an isolated 
quantum system:
\begin{equation}
\frac{\partial q_t}{\partial t} + v_H \cdot \nabla q_t =0 \nonumber
\end{equation}

This is a convection equation, typically used to describe an incompressible fluid with zero mass diffusivity, 
as dilute gases at very low temperatures. We will come back to this point later, when discussing the physical 
interpretation of our results.



\paragraph*{Liouville's theorem for GQM.} As a further point of contact with the techniques of classical 
statistical mechanics and kinetic theory, we now show that a generic Hamiltonian dynamics for the 
geometric quantum state satisfies Liouville's theorem \cite{Soto2017KineticTheoryTransport}. Indeed, by writing explicitly the total 
derivative of $q_t(Z)$ with respect to time, inserting Hamilton's equations of motion (Eq.~\eqref{eq:HAM_EOM}) and then using the continuity equation we get
\begin{align*}
\frac{d q_t}{d t} &= \sum_n \frac{\partial q_t}{\partial p^n} \frac{dp^n}{dt}+\frac{\partial q_t}{\partial \phi^n} \frac{d\phi^n}{dt} + \frac{\partial q_t}{\partial t} \nonumber\\
& = \left\{ q_t, E\right\} - \nabla \cdot q_t v_H \nonumber \\
& = 0
\end{align*}
Thus, the probability density is constant along every Hamiltonian trajectory in quantum state space. Equivalently, the Hamiltonian flow preserves the Fubini–Study volume and transports the geometric quantum state incompressibly: an infinitesimal region of $\mathcal{P}(\mathcal{H})$ may be displaced and deformed by the dynamics, but its Fubini–Study volume and the probability mass it carries are preserved. For a discrete ensemble this is the continuum counterpart of the fact that every carrier keeps its probability weight $x_\alpha$ while all carriers are propagated by the same unitary flow.

This result establishes the isolated-system baseline for the non-equilibrium dynamics considered below. Hamiltonian evolution can rearrange an ensemble across quantum state space, but it cannot by itself create local compression or dilation of probability with respect to the invariant Fubini–Study measure. Such effects require additional terms in the effective ensemble dynamics—for example branch-weight redistribution, dissipation, or diffusion generated by the coupling to an environment. In this sense, deviations from Liouvillian transport provide a direct characterization of the mechanisms through which a projected ensemble can approach equilibrium or a non-equilibrium stationary state.

\section{Kinetic theory of projected ensembles}
\label{sec:DYN}

As we move to analyze how open quantum systems scramble probability around the
quantum state space, the goal of this section is to provide a microscopic approach to the transport of probability: a kinetic theory of how the probability mass of finding our system  in a region of its quantum state space changes as a result of its interaction with a structured, non-thermal, environment. The main outcome of this section is a concrete set of microscopic equations that determines
the evolution of the geometric quantum state of an open quantum system interacting with a finite or countably infinite environment.

In order to work with concrete expressions, we work in this section with a particular type of ensemble, namely the \textit{projected ensemble}\cite{ippolitiDynamicalPurificationEmergence2023, goldsteinDistributionWaveFunction2006, Anza2021DensityMatricesGeometric}. This ensemble arises when the system of interest $S$ (with Hilbert space $\mathcal{H}_S$) is entangled with an environment $E$ (with Hilbert space $\mathcal{H}_E$) in a global pure state $\ket{\psi_{SE}}$. Expanding on a fixed environmental basis $\{\ket{e_\alpha}\}_{\alpha=0}^{D_E-1}$,
\begin{equation}
    \ket{\psi_{SE}} = \sum_{\alpha=0}^{D_E-1}\sqrt{x_\alpha}\,\ket{\chi_\alpha}\ket{e_\alpha}\,, \nonumber
\end{equation}
with $\ket{\chi_\alpha} \in \mathcal{H}_S$ identifying the conditional pure state left on the system when the environment is found in $\ket{e_\alpha}$, and $x_\alpha = \bra{e_\alpha}\rho^E\ket{e_\alpha}$ (with $\rho^E$ being the reduced environmental density matrix) identifying the probability of measuring the environment in the $\ket{e_\alpha}$ state. The pair $\{x_\alpha, \Gamma_\alpha = Z(\chi_\alpha)\}$ is exactly the data specifying a GQS of the form \eqref{eq:gqs_opn}, so the projected ensemble is the GQS of the system relative to a choice of conditional basis on the environment.

Calling $H_S$ and $H_E$ the Hamiltonian operators of the system and environment, respectively, 
the total Hamiltonian of the joint system is $H = H_S + H_E + H_{\mathrm{int}}$. Since $H_{\mathrm{int}}$
is the interaction term between system and environment, we can always put it in the form
\begin{equation}
H_{\mathrm{int}} = \sum_{k=1}^M A^{(k)} \otimes B^{(k)}~, \nonumber
\end{equation}
where $A^{(k)}$ and $B^{(k)}$ are operators with support on $\mathcal{H}_S$ and $\mathcal{H}_E$,
respectively. With a slight abuse of notation we will often conflate $A^{(k)}$ with $A^{(k)}\otimes \mathbb{I}_E$
and $B^{(k)}$ with $\mathbb{I}_S \otimes B^{(k)}$, where $\mathbb{I}_S$ and $\mathbb{I}_E$ are, respectively,
the identity operator on $\mathcal{H}_{S}$ and $\mathcal{H}_E$.
Here we do not impose a specific choice for the basis of the system and environment. While a natural one
is the bases that diagonalize the non-interacting Hamiltonians $H_S$ and $H_E$, there are cases
where a different choice is more appropriate. For example, in the case of a spin-$1/2$ chain one might be interested 
in using the computational basis. Thus, we keep things general and use a generic basis, with no particular 
properties with respect to the algebra of observables: $\left\{ \ket{a_j}\right\}_{j=0}^{d_S-1}$ and $\left\{ \ket{e_\alpha}\right\}_{\alpha=0}^{d_E-1}$.

Since the goal is to derive a dynamic equation for $\left\{x_\alpha(t) \right\}$ and $\left\{\Gamma_\alpha(t)\right\}$,
we begin with the overall Schroedinger equation for the pure state $\ket{\psi(t)}$ of the joint system+environment, written
in the generic tensor product basis defined above, with $\psi_{j\alpha}(t) = \braket{a_j,e_\alpha}{\psi(t)}$:
\begin{equation}
i\hbar \frac{d \psi_{j\alpha}(t)}{dt} = \sum_{k,\beta} H_{j\alpha;k\beta} \psi_{k\beta}(t)~, \nonumber
\end{equation}
where
\begin{align*}
H_{j\alpha;k\beta} &\coloneqq \bra{a_j,e_\alpha} H \ket{a_k,e_\beta}\\
&= \left(H_S\right)_{jk}\delta_{\alpha \beta} + \delta_{jk} \left(H_E\right)_{\alpha \beta} + \sum_{n=1}^M A_{jk}^{(n)} B^{(n)}_{\alpha \beta}~.\nonumber 
\end{align*}
We now collect the probability $x_\alpha(t)$ and the states $\ket{\chi_\alpha(t)}$ into
a single quantity: a non-normalized ket $\ket{\Phi_\alpha(t)} \coloneqq \sqrt{x_\alpha}\ket{\chi_\alpha} \in \mathcal{H}_S$.
By plugging Schroedinger's equation into the time-derivative of $\ket{\Phi_\alpha(t)}$ we get the following set of $d_E$
coupled linear equations:
\begin{equation}
i\hbar \frac{d \ket{\Phi_\alpha}}{dt} = H_S \ket{\Phi_\alpha} + \sum_{\beta} \left[\left( H_E\right)_{\alpha\beta}\ket{\Phi_\beta} + \hat{M}_{\alpha \beta}\ket{\Phi_\beta}\right]~, \nonumber
\end{equation}
where $\hat{M}_{\alpha\beta}$ is a set of operators acting on the system, defined as 
\begin{equation}
\hat{M}_{\alpha \beta} \coloneqq \sum_{k=1}^M B^{(k)}_{\alpha\beta} A^{(k)} \nonumber
\end{equation}
This can be further manipulated to separate the non-interacting part, acting on each vector $\ket{\Phi_\alpha}$, 
from the interacting part, acting on $\ket{\Phi_\beta}\neq \ket{\Phi_\alpha}$. The final form of our kinetic equation is
\begin{equation}
i\hbar \frac{d \ket{\Phi_\alpha}}{dt} = \hat{H}_\alpha \ket{\Phi_\alpha} + \sum_{\beta \neq \alpha} \hat{V}_{\alpha \beta}\ket{\Phi_\beta}~,\label{eq:kinetic}
\end{equation}
where $\hat{H}_\alpha$ is the single-particle Hamiltonian 
\begin{equation}
\hat{H}_\alpha \coloneqq H_S + \left( H_E\right)_{\alpha \alpha} \mathbb{I}_S + \sum_{k=1}^M B^{(k)}_{\alpha \alpha}A^{(k)}~,\label{ed:def_Halpha}
\end{equation}
and the interaction between particles is mediated by the set of operators $\hat{V}_{\alpha \beta}$
\begin{equation}
\hat{V}_{\alpha \beta} \coloneqq \left( H_E\right)_{\alpha \beta} \mathbb{I}_S + \sum_{k=1}^M B^{(k)}_{\alpha \beta} A^{(k)}~.\label{ed:def_V}
\end{equation}
The remaining step is to connect this with the quantities that determine the geometric quantum state of the system: $x_\alpha$ and $\Gamma_\alpha$.
This is easily done by using the original definition $\ket{\Phi_\alpha} = \sqrt{x_\alpha} \ket{\chi_\alpha}$. Since $\Gamma_\alpha(t) \leftrightarrow \ket{\chi_\alpha}$,
we have $\Gamma^j_\alpha(t) = \frac{\braket{a_j}{\Phi_\alpha(t)}}{\sqrt{x_\alpha(t)}}$ and $x_\alpha(t) = \braket{\Phi_\alpha(t)}{\Phi_\alpha(t)}$. 



Thus, Equation~\eqref{eq:kinetic} provides a way to describe the dynamics of an open 
quantum system as a set of $d_E$ linear, coupled, differential equations, for the 
geometric quantum state $\mu_t$: they provide evolution equations for $x_\alpha$ and $\Gamma_\alpha$
which, by definition, satisfy the continuity equation in Eq.~\eqref{eq:continuity}. By numerically
simulating the globally unitary evolution, we can therefore access and study both
phenomenological terms in the continuity equation: the probability flux $J_t$ and
the sink/source term $\sigma_t$.

It is worth noting that the emergence of this set of coupled equations reinforces the 
interpretation of $\Gamma_\alpha(t)$ as behaving as ``particle-like'', with their own local 
notion of energy, represented by $H_\alpha$ and their pair-wise interaction $\hat{V}_{\alpha \beta}$. 
While not treated here, by using the approach developed in Ref.~\cite{Anza2021DensityMatricesGeometric} these results can be extended to include finite-systems interacting with infinite-dimensional environments. A similar approach, grounded in the Feshbach projection operator technique, was taken by Gaspard and Nagaoka in~\cite{gaspardNonMarkovianStochasticSchrodinger1999}. They ground their work in a perturbative expansion in the coupling parameter, which then leads to a stochastic non-Markovian Schroedinger equation limited to weak coupling situations. Here we avoid this route and leverage the geometric approach to go beyond weak coupling regimes.

\subsection{Effective equations for simple interactions}\label{subsec:OQS}

In order to derive explicit forms of the source and flux terms, we consider the following simplification. Let the system and environment interact with $H_{\mathrm{int}} = A \otimes B$, then the $\hat{M}_{\alpha\beta}$ have only a single contributing term such that the single particle Hamiltonian in Eq.~\eqref{ed:def_Halpha} and the interaction operator in Eq.~\eqref{ed:def_V} have no sum over $k$. By choosing an environmental basis which diagonalizes $B$, the kinetic equation (Eq. \eqref{eq:kinetic}) allows the time evolution of $x_\alpha(t)$ to be expressed analytically as:
\begin{equation}
    \dot{x}_\alpha = -\frac{i}{\hbar}[H_E,\rho^E]_{\alpha\alpha} \nonumber
\end{equation}
where $H_E$ is the environmental Hamiltonian and $\rho^E = \mathrm{Tr}_S(\ket{\Psi_{SE}}\bra{\Psi_{SE}})$ with elements $(\rho^E)_{\alpha\beta} = \braket{\Phi_\beta}{\Phi_\alpha}$. This means that the probability mass carried by a single point changes only insofar as the environmental reduced density matrix fails to commute with the environmental Hamiltonian. Likewise, we can write an explicit expression for the evolution of $\Gamma^j_\alpha(t)$:
\begin{multline}
    \dot{\Gamma}_\alpha^j = \frac{i}{\hbar}\left(\left\{\frac{\dot{x}_\alpha}{2x_\alpha} - (H_E)_{\alpha\alpha}\right\}\Gamma^j_\alpha \right. \\ \left. - \sum_k((H_S)_{jk} + b_\alpha A_{jk})\Gamma_\alpha^k \right. \\ \left. - \sum_{\beta\ne\alpha}\sqrt{\frac{x_\beta}{x_\alpha}}(H_E)_{\alpha\beta}\Gamma^j_\beta\right) \nonumber
\end{multline}
where $A_{jk}=\bra{a_j}A\ket{a_k}$ and $b_\alpha$ are the eigenvalues of $B$. Now, using these evolution equations we can directly compute the source and flux terms: 
\begin{equation}
    \sigma_t = -\frac{i}{\hbar}\sum_\alpha [H_E,\rho^E]_{\alpha\alpha}\delta_{\Gamma_\alpha} \nonumber
\end{equation}
\begin{multline}
    J_t^j = \frac{i}{\hbar}\sum_\alpha\delta_{\Gamma_\alpha}\left( \left\{\frac{\dot{x}_\alpha}{2} - x_\alpha (H_E)_{\alpha\alpha}\right\}\Gamma^j_\alpha \right. \\ \left. - x_\alpha\sum_k((H_S)_{jk} + b_\alpha A_{jk})\Gamma^k_\alpha \right. \\ \left. - \sum_{\beta\ne\alpha} \sqrt{x_\alpha x_\beta}(H_E)_{\alpha\beta}\Gamma^j_\beta \right) \nonumber
 \end{multline}

So far, we have derived the microscopic, kinetic, equations regulating the 
probability transport across the quantum state space. In the next few sections, using
both analytical and numerical approaches, we explore the theoretical framework developed
so far to look at the phenomenology of probability transport in concrete physical systems.

\section{Qubit in a Caldeira-Leggett environment}
\label{sec:EXAMPLES2}

As an example of the usefulness of this framework, we show that a phenomenological description of dissipative dynamics arises naturally from the continuity equation for stochastic dynamics. To this end, we consider as our system a qubit interacting with an environment modeled as a Caldeira--Leggett bath~\cite{caldeiraQuantumTunnellingDissipative1983, caldeiraPathIntegralApproach1983, leggettDynamicsDissipativeTwostate1987}, traditionally realized as an infinite collection of non-interacting harmonic oscillators, coupled linearly to the system, with Ohmic correlation functions. When the coupling is weak and the bath is large and unstructured, its degrees of freedom can be integrated out, leaving a Langevin equation \cite{ford_quantum_1988,leggettDynamicsDissipativeTwostate1987, gardiner_quantum_2004} for the system alone with three components: a deterministic drift generated by a renormalized system Hamiltonian, a friction linear in the velocities, and a Gaussian white-noise force whose strength is set by the bath temperature. For a two-level system with effective system Hamiltonian $H_S = \delta\,\sigma^x + \epsilon\,\sigma^z$, the energy function on $\mathbb{C}P^{1}$ in canonical coordinates (probabilities and phases) is
\begin{equation}
    E(p,\phi) = 2\delta\sqrt{p(1-p)}\,\cos\phi + \epsilon\,(1-2p)\,,
    \label{eq:langevin-energy}
\end{equation}
and the dynamics are governed by the stochastic differential equations
\begin{align*}
    \dot{p} & = \partial_\phi E + V_p + W_p, \\
    \dot{\phi} & = -\partial_pE + V_\phi + W_\phi\,,
\end{align*}
where $V_p$ and $V_\phi$ model dissipation via friction or decay, while $W_p$ and $W_\phi$ are stochastic variables modeling the system-environment interaction. For a large, unstructured environment these stochastic terms are well described by a Gaussian process, with
$\mathbb{E}[W_a(s+t)W_b(s)] = \mathbb{E}[W_a(t)W_b(0)]\approx \delta_{ab}\gamma_a\delta(t)$,
$a,b\in\{p,\phi\}$, and $\gamma_a\propto k_BT$~\cite{Anza2022GeometricQuantumThermodynamics}. For the purposes of this discussion we consider a minimal dissipative model, with friction acting only in the probability channel, $V_p = k_f\dot{\phi}$ and $V_\phi = 0$, and noise likewise injected only in the probability channel, $W_\phi = 0$. The Langevin equations governing the dynamics of our ensemble are then
\begin{subequations}
\label{eq:langevin}
\begin{align}
    \dot{p} & = \partial_\phi E - k_f\,\partial_p E + \sqrt{\gamma}\,\xi(t)\,, \\
    \dot{\phi} & = -\partial_p E \,,
\end{align}
\end{subequations}
or, written explicitly using the functional form of the energy in Eq.~\eqref{eq:langevin-energy},
\begin{align*}
    \dot{p} & = -2\delta\sqrt{p(1-p)}\,\sin\phi + k_f\dot{\phi} + \sqrt{\gamma}\,\xi(t)\,, \\
    \dot{\phi} & = -\delta\,\frac{1-2p}{\sqrt{p(1-p)}}\,\cos\phi + 2\epsilon\,.
\end{align*}

The probability--phase coordinates describe the principal chart $0<p<1$ with $\phi\sim\phi+2\pi$. The endpoints $p=0$ and $p=1$ correspond to the two poles of the Bloch sphere, where the phase coordinate becomes physically irrelevant: all values of $\phi$ at a given endpoint represent the same point of $\mathbb{C}P^1$. We complete the effective stochastic dynamics by imposing periodicity in $\phi$ and reflecting boundary conditions at $p=0,1$. These conditions preserve the physical interval $0\leq p\leq1$ and conserve total probability. They are part of the effective Langevin model obtained after the
Caldeira--Leggett reduction and are inherited by the corresponding Fokker--Planck equation.

\begin{figure*}[t!]
\centering
\includegraphics[width=0.95\textwidth]{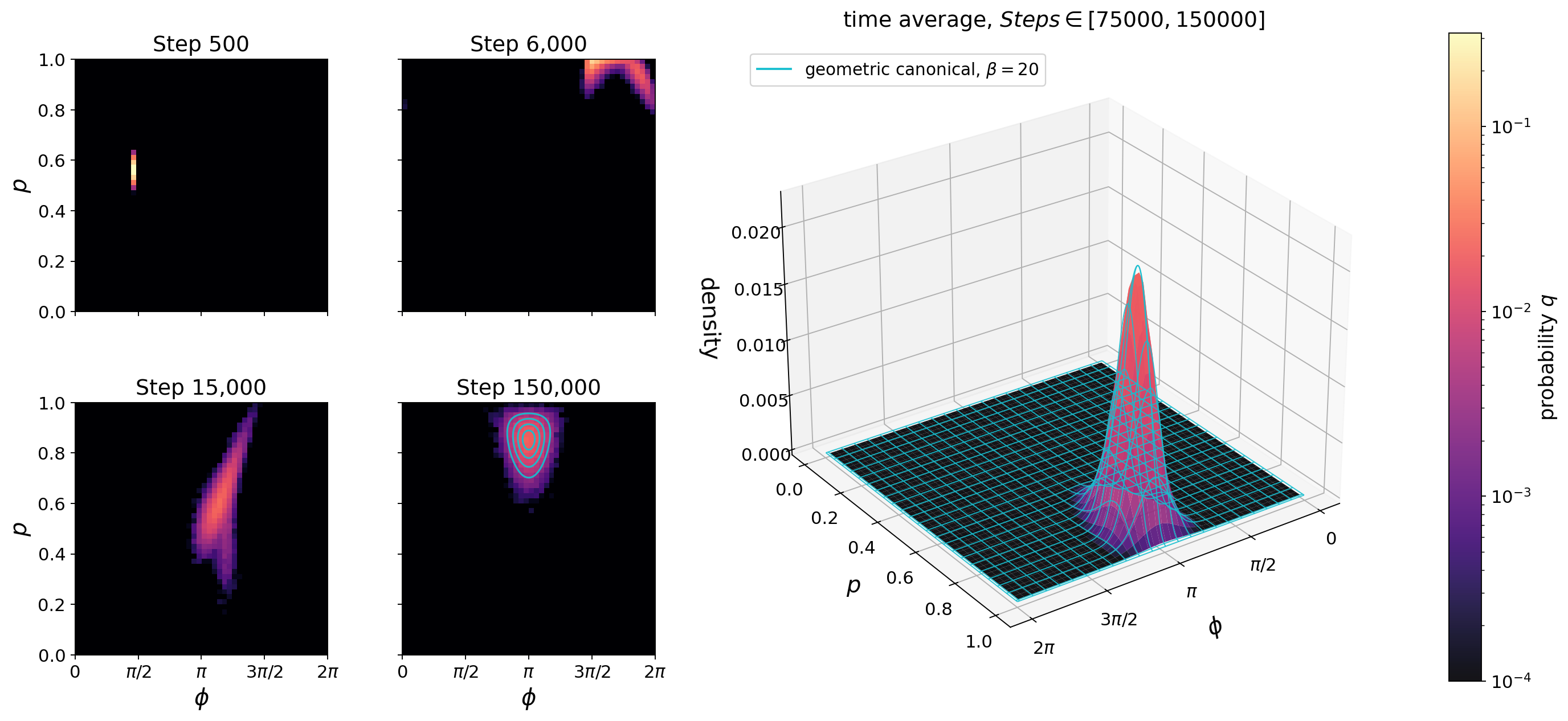}
\caption{Left: Relaxation of a single stochastic trajectory of the qubit–Caldeira–Leggett Langevin equations, Eq.~\eqref{eq:langevin}, toward the geometric canonical ensemble. Each panel shows the time-averaged histogram of $(p,\phi)$ accumulated over the trajectory up to the indicated step, exploiting the ergodicity of the process to approximate the ensemble $q_t$ from a single realization of the noise rather than an average over many. Starting from a narrow initial distribution, the trajectory is driven by the Hamiltonian drift toward the pole $p=1$ before the dissipative and diffusive terms relax it onto a stationary profile. Right: Time-averaged ensemble from step 75,000 to step 150,000. By step 150,000, the accumulated histogram has converged onto the predicted stationary state $q_\mathrm{eq} \propto e^{-\beta E(p,\phi)}$ of Eq.~\eqref{eq:canonical-ensemble} (cyan contours, $\beta = 2k_f/\gamma = 20$), confirming the Fokker-Planck prediction numerically.}
\label{fig:gqs_dynamics2}
\end{figure*}

Provided the resulting Markov process is ergodic, stationary ensemble statistics can be estimated from the aggregated time series of a single sufficiently
long trajectory~\cite{Anza2022GeometricQuantumThermodynamics}. By numerically implementing the differential equations above, we can directly simulate the dynamics of the system, with the resulting ensemble being shown in Figure \ref{fig:gqs_dynamics2}.

Equations~\eqref{eq:langevin} describe one realization of the noise, that is, one member of the ensemble. The object of interest in the present framework is not the individual trajectory but the distribution $q_t(p,\phi)$ that the trajectories collectively realize on $\mathbb{C}P^1$. To obtain an evolution equation for the ensemble itself, we use the Langevin equations for $\dot p$ and $\dot \phi$ to derive a Fokker--Planck equation for $q_t(p,\phi)$. We first introduce the vector notation $\mathbf{x} = (p, \phi)$, $\mathbf{A} = (\partial_\phi E-k_f\partial_p E,\, -\partial_p E)$, and $\mathbf{B} = (\sqrt{\gamma},\, 0)$, so that Eqs.~\eqref{eq:langevin} reduce to a single stochastic differential equation (SDE),
\begin{equation}
    \dot{\mathbf{x}} = \mathbf{A} + \mathbf{B}\,\xi(t)\,.\nonumber
\end{equation}
Integrating this expression over an infinitesimal timestep $dt$ gives the standard form of an SDE describing an It\^o process driven by Brownian motion,
\begin{equation}
    d\mathbf{x} = \mathbf{A}\,dt + \mathbf{B}\,dW\,. \label{eq:general_langevin}
\end{equation}

To derive the Fokker–Planck evolution of the ensemble, we introduce an arbitrary twice-differentiable test function $f(\mathbf{x})=f(p,\phi)$
and determine the infinitesimal generator $\mathcal{L}$ of the Ito process. The generator governs the evolution of the expectation value of $f$ through the backward Kolmogorov equation
\begin{equation}
\int_{\mathbb{C}P^1}\!\!\!\!\!
f\partial_t q_tdV_{\mathrm{FS}}=
\int_{\mathbb{C}P^1}\!\!\!\!\!
(\mathcal{L}f)q_tdV_{\mathrm{FS}}. \nonumber
\end{equation}
The Fokker–Planck equation is simply the resulting evolution of $q_t$, generated by the adjoint $\mathcal{L}^{\dagger}$:
\begin{equation}
\partial_tq_t=\mathcal{L}^{\dagger}q_t. \nonumber
\end{equation}
The explicit derivation of $\mathcal{L}$ and $\mathcal{L}^\dagger$ is given in Appendix \ref{app:fp-qubit}, which result in the following Fokker--Planck equation:
\begin{equation}
\frac{\partial q_t}{\partial t} = - \Poiss{q_t}{E} + k_f\frac{\partial}{\partial p}
\left( q_t\frac{\partial E}{\partial p} \right) + \frac{\gamma}{2} \frac{\partial^2q_t}{\partial p^2}~.
\label{eq:qubit-fokker-planck}
\end{equation}

By direct inspection, we can see that this can be brought in the form of the continuity equation, with $\sigma_t=0$ and $J_t = J_{\mathrm{Ham}}+J_{\mathrm{Fric}}+J_{\mathrm{Diff}}$:
\begin{subequations}
\begin{align*}
    &J_{\mathrm{Ham}} = q_t v_H= q_t (\partial_\phi E,-\partial_pE)~,\\
    &J_{\mathrm{Fric}} = (-k_fq_t \partial_pE,0)~,\\
    &J_{\mathrm{Diff}} = (-\frac{\gamma}{2} \partial_p q_t, 0)~.
\end{align*}
\end{subequations}
The three different elements of the dynamics, already present in the stochastic differential equation formulation, determine the evolution of the density $q_t$ by generating different types of probability fluxes. We can see the Hamiltonian flux $J_{\mathrm{Ham}}$, with the Poisson brackets, generated by the Hamiltonian vector field $v_H$; the dissipative flux generated by the friction term $J_{\mathrm{Fric}}$, which pushes the system towards its ground state; and the diffusion term $J_{\mathrm{Diff}}$, proportional to the gradient of the density, generating an osmosis of probability density that tends to make the distribution uniform along the $p$ coordinate.

Finally, we show that the canonical ensemble is a particular stationary state of the dynamics, in which the friction and diffusion fluxes balance each other. Consider the ansatz $q_t = f(E)$ for a stationary state, describing an ensemble that depends on the state only through its energy. First we remember that the Hamiltonian flux has no divergence $\nabla \cdot J_{\mathrm{Ham}}= -\Poiss{f(E)}{E} = 0$ -- in agreement with Liouville's theorem proven in Section \ref{sec:IQS}. Then when $J_{\mathrm{Fric}}+J_{\mathrm{Diff}}=0$ we obtain a stationary state because $\partial_t q_t =- \nabla \cdot (J_{\mathrm{Fric}}+J_{\mathrm{Diff}})=0$. The equilibrium condition leads to 
\begin{align}
    -k_ff\,\partial_pE - \frac{\gamma}{2}\partial_p f &= -\left[ k_f f + \frac{\gamma}{2}f' \right]\partial_p E=0, \nonumber
\end{align}
which is solved when $f(E)\propto e^{-\beta E}$, provided that $\beta=\frac{2k_f}{\gamma}$. This provides analytical proof that the canonical form of the projected ensemble 
\begin{equation}
    q_\mathrm{eq}(p,\phi) = \frac{1}{Z_\beta}\,e^{-\beta E(p,\phi)}\,, \qquad \beta = \frac{2k_f}{\gamma}\,, \label{eq:canonical-ensemble}
\end{equation}
with $Z_\beta = \int_{\mathbb{C}P^1}e^{-\beta E(p,\phi)}\, dV_{FS}$ the geometric partition function, is the equilibrium solution for a qubit in a Caldeira-Leggett environment. This result is in direct agreement with the numerical study performed in~\cite{Anza2022GeometricQuantumThermodynamics} as well as with the numerics presented in Figure \ref{fig:gqs_dynamics2} with mean squared error between the simulated ensemble and the theoretically predicted canonical distribution $\Delta_{MSE} = 1.92\times10^{-8}$.

\section{Qubit in a spin-star environment}
\label{sec:SPINSTAR}
We now look at the example of a single qubit of interest interacting with $N$ environmental spins equally spaced from the system qubit that do not interact with one another. This has been studied in the literature as the ``spin-star model"~\cite{hamdouniExactlySolvableModel2009, yuanDynamicsDrivenSpin2011, bortzExactDynamicsInhomogeneous2007, mahdianExactDynamicsSingleQubit2015, hamdouniExactDynamicsTwoqubit2006, breuerNonMarkovianDynamicsSpin2004, wangNonMarkovianDynamicsSpin2013, dengQuantumPhaseTransitions2008, korkmazQuantumtoclassicalTransitionSpin2022}. The Hamiltonian we consider is of the form
\begin{equation}
    H = \frac{\omega}{2}\sigma_s^x + \Omega\sigma_s^z\otimes\frac{1}{N}\sum_{k=1}^N\sigma_k^z \label{eq:spin-star}
\end{equation}
describing a central spin $s$ in a transverse field of strength $\omega$, Ising coupled ($ZZ$) to each of the $N$ environmental spins with uniform strength $\Omega/N$~\cite{wangNonMarkovianDynamicsSpin2013}. Two features make \eqref{eq:spin-star} a natural first analytical test case. The interaction couples the system only to the \emph{collective} bath operator $\tfrac{1}{N}\sum_k\sigma_k^z$, so the dynamics is invariant under permutations of the environmental spins; and the model carries no bath self-Hamiltonian, a choice we make deliberately to force the source term to vanish, as shown below. Despite this simplification, as we see now the projected ensemble has a non-trivial dynamics, which can be studied analytically using the tools developed in Sections \ref{sec:DYN} and \ref{sec:IT}.

We use the bipartite decomposition of the global pure state into conditional branches,
\begin{equation}
    \ket{\Psi_{SE}} = \sum_{\alpha}\sqrt{x_\alpha}\,\ket{\chi_\alpha}\ket{e_\alpha} , \qquad \ket{\Phi_\alpha} \equiv \sqrt{x_\alpha}\,\ket{\chi_\alpha} \nonumber
\end{equation}
where $\ket{\Phi_\alpha}$ is the unnormalized system state conditioned on the environment occupying $\ket{e_\alpha}$, and $x_\alpha = \braket{\Phi_\alpha}{\Phi_\alpha} = \bra{e_\alpha}\rho^E\ket{e_\alpha}$ is the corresponding branch weight. Matching \eqref{eq:spin-star} to the generic interacting form $H = H_S + H_E + A^S\otimes B^E$ gives
\begin{align*}
    H_S & = \tfrac{\omega}{2}\sigma_s^x, \qquad H_E = 0, \\ A^S & = \Omega\,\sigma_s^z, \qquad B^E = \tfrac{1}{N}\textstyle\sum_k\sigma_k^z
\end{align*}
We choose the conditional basis $\{ \ket{e_\alpha} \}$ to diagonalize the interaction $B^E$, which for \eqref{eq:spin-star} is the computational basis. Two consequences follow immediately. First, from the microscopic expression for the branch-weight dynamics derived in Sec.~\ref{subsec:OQS},
\begin{equation}
    \dot{x}_\alpha = -\tfrac{i}{\hbar}[H_E, \rho^E]_{\alpha\alpha} = 0 \qquad \longrightarrow \quad \sigma = 0 \label{eq:sigma-zero}
\end{equation}
Physically, every $\sigma_k^z$ commutes with \eqref{eq:spin-star}, so the bath is frozen in the conditional basis: the populations $x_\alpha$ are constants of motion and no probability weight is exchanged between branches. Second, projecting the Schr\"odinger equation onto $\{ \ket{e_\alpha} \}$ and writing $s_k^{(\alpha)} = \pm1$ for the eigenvalues of $\sigma_k^z$,
\begin{align}
    i\hbar\,\partial_t \ket{\Phi_\alpha} & = \left( \frac{\omega}{2}\sigma_s^x + \Omega\,\mu_\alpha\,\sigma_s^z \right)\ket{\Phi_\alpha} \equiv H^\alpha_{\mathrm{eff}}\,\ket{\Phi_\alpha},\label{eq:heff} \\
    \mu_\alpha & \equiv \frac{1}{N}\sum_{k=1}^N s_k^{(\alpha)} \nonumber
\end{align}
so each branch evolves as a \emph{closed} effective single qubit, driven by a fixed Hamiltonian set by the net bath magnetization $\mu_\alpha$ of that branch.

\subsection{Branch stationary states and their superposition}

Because every branch \eqref{eq:heff} is a closed single-qubit system, and the source term vanishes by \eqref{eq:sigma-zero}, the two conditions leading to \eqref{eq:stat-energy} are met branch by branch: the stationary distribution of branch $\alpha$ is an arbitrary function of the conserved branch energy, $q_\alpha = f(E_\alpha)$. Identifying \eqref{eq:heff} with the generic single-qubit decomposition, the energy function reads
\begin{equation}
    E_\alpha(p,\phi) = \omega\sqrt{p(1-p)}\,\cos\phi + \Omega\mu_\alpha(1-2p), \nonumber
\end{equation}
Since the branch weights are conserved by \eqref{eq:sigma-zero}, the full stationary distribution is a \emph{fixed} convex combination of the branch solutions,
\begin{equation}
    q(p,\phi) = \sum_\alpha x_\alpha\,q_\alpha(p,\phi), \label{eq:q-convex}
\end{equation}
It remains to fix the weights through an initialization of the bath. Preparing each environmental spin independently in the coherent pure state of mean magnetization $m_0$, $\ket{\psi} = \sqrt{\frac{1+m_0}{2}}\,\ket{0} + \sqrt{\frac{1-m_0}{2}}\,\ket{1}$, the product structure $\rho^E = \bigotimes_k\rho^{(k)}$ gives a branch weight that depends only on the number $n$ of up-spins in the bitstring $\alpha$,
\begin{equation}
    x_\alpha = \left(\frac{1+m_0}{2}\right)^{\!n}\left(\frac{1-m_0}{2}\right)^{\!N-n}.\nonumber
\end{equation}
The permutation symmetry of \eqref{eq:spin-star} now collapses the sum over the $2^N$ branches in \eqref{eq:q-convex} into a sum over the $N+1$ distinct magnetization sectors, weighted by their multiplicities $\binom{N}{n}$:
\begin{multline}
    q_{st}(p,\phi) = \sum_{n=0}^N\binom{N}{n}\left( \frac{1+m_0}{2} \right)^{\!n}\left(\frac{1-m_0}{2}\right)^{\!N-n} \\ f\!\left( \omega\sqrt{p(1-p)}\,\cos\phi + \Omega\mu_n(1-2p)\right). \label{eq:q-binomial}
\end{multline}
with $\mu_n = 2n/N-1$. Equation~\eqref{eq:q-binomial} holds for any $N$; in the thermodynamic limit the binomial weight concentrates and the sum may be replaced by a Gaussian integral over $\mu$ via the central limit theorem.

\begin{figure*}
    \centering
    \includegraphics[width=0.95\textwidth]{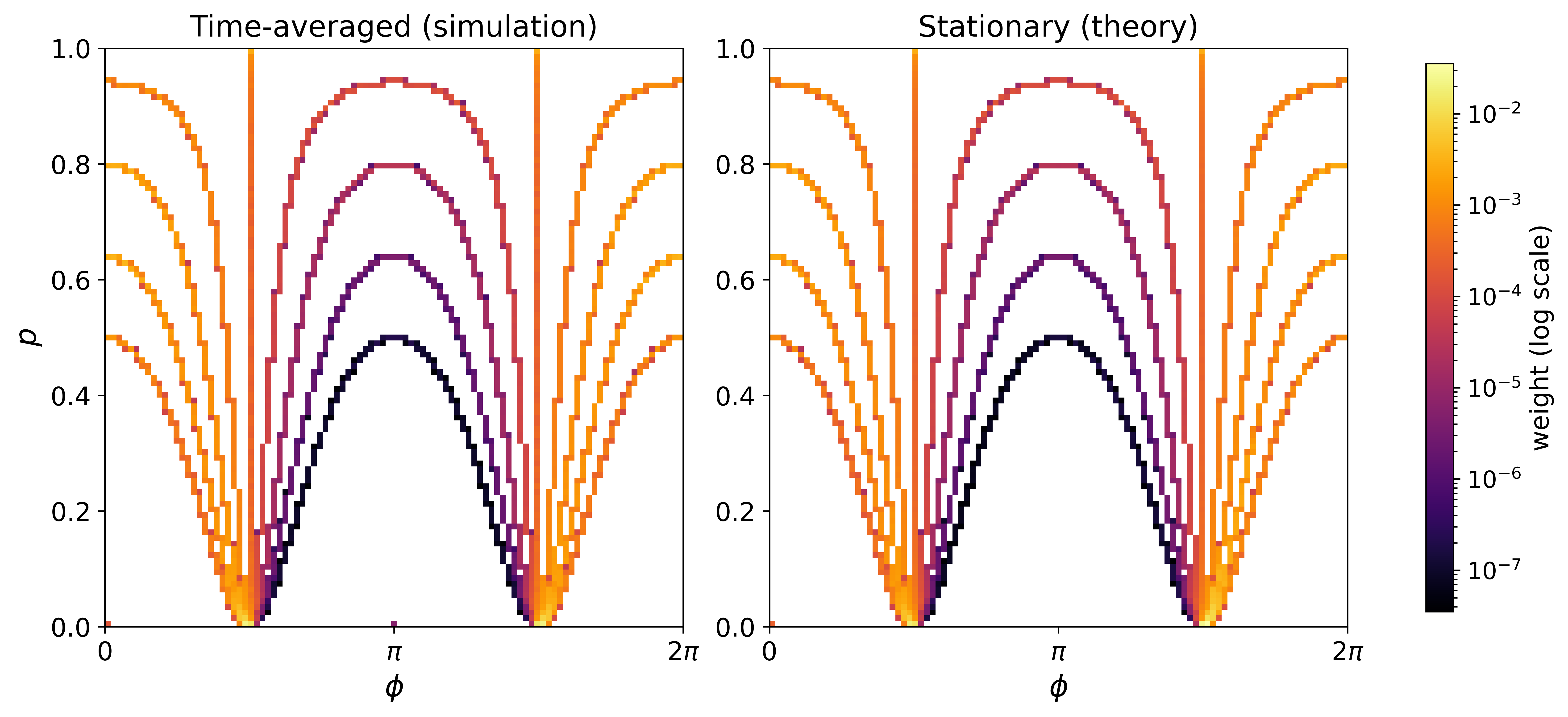}
    \caption{Time-averaged projected ensemble obtained by exact diagonalization of
Eq.~\eqref{eq:spin-star} (left), compared with the analytic stationary state of
Eq.~\eqref{eq:q-binomial} (right), over the $(p,\phi)$ plane. The central spin is
initialized in $\ket{0}$ and the bath is prepared at mean magnetization $m_0 = 0.5$.
Weights are shown on a logarithmic color scale. The two distributions agree to
$\Delta_{\mathrm{MSE}} = 1.94\times 10^{-9}$, at the level of the sampling noise,
confirming the prediction.}
    \label{fig:spin-star-numerics}
\end{figure*}

\subsection{Numerical Verification}

The functional form of $f$ in \eqref{eq:q-binomial} is fixed only once the initial system state is specified. For a pure initial state $q(p,\phi,0) = \delta(p-p_0)\delta(\phi-\phi_0)$, each branch is confined to the level set $E_n(p,\phi) = E_n(p_0,\phi_0)$, so that
\begin{multline}
    f\left( E_n(p,\phi) \right) = \delta\left( \omega\left[ \sqrt{p(1-p)}\,\cos\phi  \right. \right. \\ \left. \left. - \sqrt{p_0(1-p_0)}\,\cos\phi_0 \right] + 2\Omega\mu_n(p_0-p) \right) \nonumber
\end{multline}
Simulating \eqref{eq:spin-star} via exact diagonalization with the central spin initialized in $\ket{0}$ and the bath prepared at mean magnetization $m_0$, we compare the analytic prediction \eqref{eq:q-binomial} against the time-averaged projected ensemble. As shown in Fig.~\ref{fig:spin-star-numerics}, the two agree closely across the $(p,\phi)$ plane, with residuals at the level of sampling noise, confirmed by the mean squared error which is computed to be $\Delta_{MSE} = 1.94\times10^{-9}$.

\section{Conclusions}
\label{sec:FINAL}

We developed a non-equilibrium theory of projected ensembles. Treated as a dynamical probability measure over the manifold of pure states, its evolution is determined by the underlying interacting dynamics of system and environment. This leads to a continuity equation on quantum state space, together with exact kinetic equations and analytical expressions for the probability flux $J_t$ and source term $\sigma_t$.

As a result, the dynamics of projected ensembles admits a classical Hamiltonian/kinetic representation on the quantum state space. For isolated systems, the probability measure is transported by the Hamiltonian flow generated by the energy function on $\mathcal{P}(\mathcal{H})$, leading directly to Liouville's theorem. For open quantum systems, the interaction with the environment enriches this structure into a kinetic theory for the probability distribution, in which the evolution is organized in terms of currents, sources and sinks, dissipation, and diffusion. Microscopically, the projected ensemble is generated by a long-range, pair-wise interacting dynamics in which the conditional states behave as carriers of probability mass. Probability is transported locally through the flux, while changes in the carrier weights redistribute it non-locally through the source term. This establishes a novel connection between projected-ensemble dynamics and classical non-equilibrium physics, thus providing a systematic framework to characterize both equilibrium and non-equilibrium stationary states.

In practice, the method of characteristics turns this transport problem into a  solvable one: the stationary continuity equation, $\nabla\cdot J=\sigma$, reduces along each characteristic curve to an ordinary differential equation for the geometric quantum state (Section \ref{sec:MOC}), giving direct analytical access to stationary ensembles that need not be thermal. We used this to solve two complementary cases in closed form, a qubit in a Caldeira--Leggett bath (Section \ref{sec:EXAMPLES2}) and a qubit in a spin-star environment (Section \ref{sec:SPINSTAR}). 

This leads to a number of forward-looking conclusions. 

First, the continuity equation together with the kinetic equations determine the entire dynamics of the projected ensemble. In particular, they also determine the  evolution of all the higher moments $\rho^{(k)}_t = \int \mu_t(dV_{\mathrm{FS}})\left( \ket{\psi}\bra{\psi}\right)^{\otimes k}$, which retain information about the underlying distribution of pure states and provide the natural observables for characterizing quantum state designs and deep thermalization. In close analogy with the BBGKY hierarchy of classical kinetic theory \cite{spohn_large_1991}, this provides a systematic route to understanding how stationary structures emerge in projected ensembles, on what time-scales and, in turn, to a new approach to random quantum-state generation.

Second, the stationary continuity equation $\nabla \cdot J = \sigma$ provides a general characterizing equation for stationary projected ensembles, well beyond the thermal regime. Deep thermalization appears as one particular possibility: the projected ensemble converges toward a universal thermal form. In the diffusive regime realized by the Caldeira–Leggett model, this equilibrium solution is selected by a fluctuation–dissipation balance between frictional and diffusive probability currents, with $q_{\mathrm{eq}}\propto e^{-\beta E}$. More generally, however, the same equation admits non-thermal solutions: the spin-star model provides a first example of this phenomenology, where the stationary ensemble preserves detailed information about the environmental branch statistics and coupling structure. 

Third, isolated unitary dynamics can transport and deform an ensemble but, by Liouville’s theorem, it cannot locally compress its probability density with respect to the Fubini–Study volume. The clustering of an ensemble around pointer states, taken to underlie the emergence of classicality \cite{touil_branching_2024,zhu_observation_2025}, must therefore originate from non-Liouvillian contributions induced by the environment or conditioning, through a compressible probability flux, redistribution of branch weights through the source term, or both.

Looking ahead, the same machinery extends naturally to the other types of conditioning discussed in the introduction, in particular monitored systems and quantum trajectories, and, since Hamiltonian flows can generically stretch and fold, it hints at a phenomenology of chaotic dynamics for projected ensembles far richer than the rigid convection of Section \ref{sec:IQS}. More broadly, these geometric tools are unraveling a fruitful connection between projected ensembles and classical non-equilibrium physics, in which one can study not only universal limiting ensembles but also the currents, stationary structures, and relaxation processes that generate them. In this sense, the present work is a significant step toward a broader non-equilibrium theory of projected ensembles, centered not only on when universal equilibrium statistics emerge, but also on the rich structures that arise when they do not.

\section*{Acknowledgments}
\label{sec:acknowledgments}

We thank James Crutchfield, David Gier, Ariadna Venegas-Li and Dhruva Karkada
for discussions on the geometric formalism of quantum mechanics and the Telluride
Science Research Center for its hospitality during visits.  This material is
based upon work supported by, or in part by, a Templeton World Charity
Foundation Power of Information Fellowship.

\newpage
\bibliography{kinetic}

\onecolumngrid
\newpage
\appendix

\section{Derivation of the Fokker-Planck equation for a qubit}\label{app:fp-qubit}

Here we derive the generator $\mathcal{L}$ of the It\^o process \eqref{eq:general_langevin}, its
adjoint $\mathcal{L}^\dagger$, and the Fokker--Planck equation \eqref{eq:qubit-fokker-planck} quoted in
Section \ref{sec:EXAMPLES2}. Throughout, $\mathbf{x}=(p,\phi)$, and $\mathbf{A}$ and $\mathbf{B}$ are the
drift and noise vectors defined there. We also use the fact that, for $D=2$, the Fubini-Study volume form
in the probability-phase chart is $dV_{FS} = dp\,d\phi/2$: since $\sqrt{g}=1/2$ is constant, integrations
by parts against $dV_{FS}$ carry no metric factors, and the covariant divergence appearing in
\eqref{eq:continuity} reduces to $\partial_p J^p_t + \partial_\phi J^\phi_t$. This is a property of the
volume element, not a statement that $\mathbb{C}P^1$ is flat.

For an arbitrary twice-differentiable test function $f(\mathbf{x}) = f(p,\phi)$, It\^o's lemma gives
\begin{equation}
df = \sum_i(\partial_{x_i}f)dx_i+
\frac{1}{2}
\sum_{ij}
(\partial_{x_i}\partial_{x_j}f)
dx_idx_j.
\label{eq:ito-lemma-local} \nonumber
\end{equation}
Using Eq.\eqref{eq:general_langevin} together with It\^o's rules
\begin{equation}
dW^2=dt,\qquad dtdW=0, \qquad dt^2=0,\nonumber
\end{equation}
this becomes
\begin{equation}
df = \left[ \sum_i A^i\partial_{x_i}f + \frac{1}{2} \sum_{ij} D^{ij} \partial_{x_i}\partial_{x_j}f \right]dt
+ \sum_i B^i\partial_{x_i}f\,dW,
\label{eq:ito-test-function}
\end{equation}
where the diffusion tensor is $D^{ij}=B^iB^j$. We identify the coefficient of $dt$ as the action of the
generator of Kolmogorov's backward equation,
\begin{equation}
\mathcal{L}f = \sum_i A^i\partial_{x_i}f + \frac{1}{2} \sum_{ij} D^{ij} \partial_{x_i}\partial_{x_j}f,
\label{eq:backward-generator}
\end{equation}
so that \eqref{eq:ito-test-function} reads compactly $df = (\mathcal{L}f)dt + \sum_i B^i\partial_{x_i}f\,dW$.
Taking the expectation value $\mathbb{E}[df]$ over the ensemble of trajectories, whose one-time marginal
density is exactly the ensemble $q_t(p,\phi)$ we wish to describe, eliminates the $dW$ term and leaves
$\frac{d}{dt}\mathbb{E}[f] = \mathbb{E}[\mathcal{L}f]$.

The Fokker--Planck equation follows by moving the derivatives in \eqref{eq:backward-generator} off the test
function and onto the density, integrating by parts once in the drift term and twice in the diffusion term:
\begin{equation}
\int_{\mathbb{C}P^1}\!\!\!\!\! (\mathcal{L}f)\,q_t\,dV_{FS} =
\int_{\mathbb{C}P^1}\!\!\!\!\! f\left[ -\sum_i\partial_{x_i}\left(A^iq_t\right)
+ \frac{1}{2}\sum_{ij}\partial_{x_i}\partial_{x_j}\left(D^{ij}q_t\right)\right]dV_{FS} + \mathcal{B},
\label{eq:adjoint-ibp}
\end{equation}

The boundary term vanishes $\mathcal{B}=0$ since the quantum state space $\mathbb{C}P^1$ has no boundaries. However, since we are working on a specific chart, in which $p\neq 0,1$, we have to make sure that the global geometry is respected. We do so by imposing the following boundary conditions:
\begin{itemize}
    \item The coordinate $\phi$ is itself cyclical. Therefore any function on the state space will be such that $f(p,\phi)=f(p,\phi+2\pi)$ since these two points are effectively one and the same
    \item At the poles $p=0,1$ the coordinate $\phi$ disappears. In order to account for the true global geometry we impose reflecting boundary conditions on $p$. As a result, there is no flux at the boundaries: $J_t^p\vert_{p=0,1}=0$
    \end{itemize}

Since $f$ is arbitrary, comparison of \eqref{eq:adjoint-ibp} with
$\int f\,\partial_tq_t\,dV_{FS} = \int(\mathcal{L}f)q_t\,dV_{FS}$ gives $\partial_tq_t = \mathcal{L}^\dagger q_t$ with
\begin{equation}
\mathcal{L}^{\dagger}q = -\sum_i\frac{\partial}{\partial x_i}\left(A^iq\right)
+ \frac{1}{2}\sum_{ij}\frac{\partial^2}{\partial x_i\partial x_j}\left(D^{ij}q\right)\,,
\label{eq:generalized-fokker-planck} \nonumber
\end{equation}
the general Fokker--Planck operator in the chart.

For the specific dynamical model considered here, $D^{ij} = \gamma\,\delta_{i,p}\delta_{j,p}$, and using the
explicit form of $\mathbf{A}$ the Fokker--Planck equation reduces to
\begin{equation}
    \frac{\partial q_t}{\partial t} = -\frac{\partial}{\partial p}\left( \left(\frac{\partial E}{\partial\phi} - k_f\frac{\partial E}{\partial p}\right) q_t \right) + \frac{\partial}{\partial\phi}\left( \frac{\partial E}{\partial p} q_t \right) + \frac{\gamma}{2}\frac{\partial^2q_t}{\partial p^2}\,.
    \label{eq:fp-explicit}
\end{equation}
The two terms that do not carry $k_f$ recombine into a Poisson bracket,
\begin{equation}
-\frac{\partial}{\partial p}\left(\frac{\partial E}{\partial \phi}q_t\right)
+\frac{\partial}{\partial \phi}\left(\frac{\partial E}{\partial p}q_t\right)
= -\Poiss{q_t}{E} - q_t\left( \frac{\partial^2E}{\partial p\,\partial\phi} - \frac{\partial^2E}{\partial\phi\,\partial p}\right)
= -\Poiss{q_t}{E}\,, \nonumber
\end{equation}
the second term vanishing by equality of mixed partials, which is the statement $\nabla\cdot v_H=0$ used in
Section \ref{sec:IQS}. Expanding the remaining derivative in \eqref{eq:fp-explicit} then returns
Eq.~\eqref{eq:qubit-fokker-planck} of the main text:

\begin{equation}
\frac{\partial q_t}{\partial t} = - \Poiss{q_t}{E} + k_f\frac{\partial}{\partial p}
\left( q_t\frac{\partial E}{\partial p} \right) + \frac{\gamma}{2} \frac{\partial^2q_t}{\partial p^2}~.\nonumber
\end{equation}

\end{document}